# A COMPARATIVE STUDY BETWEEN LINEAR AND NONLINEAR SPEECH PREDICTION

Marcos Faúndez*, Enric Monte**, Francesc Vallverdú**

*Escola Universitària Politècnica de Mataró.
**Department of Signal Theory and Communications.
Polithecnic University of Catalonia
Avda. Puig i Cadafalch 101-111 08303 MATARO( BARCELONA, SPAIN)
e-mail: marcos@gps.tsc.upc.es fax: 34-3-7570524

ABSTRACT

This paper is focused on nonlinear prediction coding, which consists on the prediction of a speech sample based on a nonlinear combination of previous samples. It is known that in the generation of the glottal pulse, the wave equation does not behave linearly [2], [10], and we model these effects by means of a nonlinear prediction of speech based on a parametric neural network model. This work is centred on the neural net weight's quantization and on the compression gain.

## 1. Introduction

Speech applications usually require the computation of a linear prediction model for the vocal tract. This model has been successfully applied during the last thirty years [1], but it has some drawbacks. Mainly, it is unable to model the nonlinearities involved in the speech production mechanism [2], and only one parameter can be fixed: the analysis order. With nonlinear models, the speech signal is better fit, and the model can be adapted more easily to the application.

Recently there is a growing interest on nonlinear prediction coding, and several contributions have appeared ([3] to [7]) on the context of neural nets [8].

This paper will cover some interesting aspects not discussed so far. Mainly they are:

- How should the network parameters be quantized for a minimal degradation ofthe predictive performance?.
- Which is the optimal representation of the network parameters?.

Some other relevant aspects such as:

- Kind of neural net (Multilayer perceptron, Elman, etc.)
- Architecture of the network (number of layers and neurons per layer, etc.)
- Training algorithm (back-propagation, Levenberg-Marquardt, etc.)
- Number of epochs, weights' initialization, etc.

## 2. Comparison between classical lpc-12 and the proposed nlpc

Just to show the performance of the NLPC scheme it is presented the following experiment: with a test database of more than 20 seconds and 8 speakers (4 male and 4 female), we have obtained the following prediction gains (Gp):

| | MALE | | FEMALE | |
|---|---|---|---|---|
| | LPC-12 | NLPC | LPC-12 | NLPC |
| AVERAGED Gp [dB] | 12.4 | 15.1 | 12.8 | 17.1 |
| )(Gp) | 5.5 | 6.4 | 6.6 | 9 |

Main selected parameters and conclusions are:

! The selected nonlinear predictor consists on a multilayer perceptron with 10 inputs (N=10), two neurons in the hidden layer with a nonlinear transfer function, and one output layer with a linear transfer function. Experimental results reveal that this size and number of neurons are a good compromise between computational requirements and system performance.

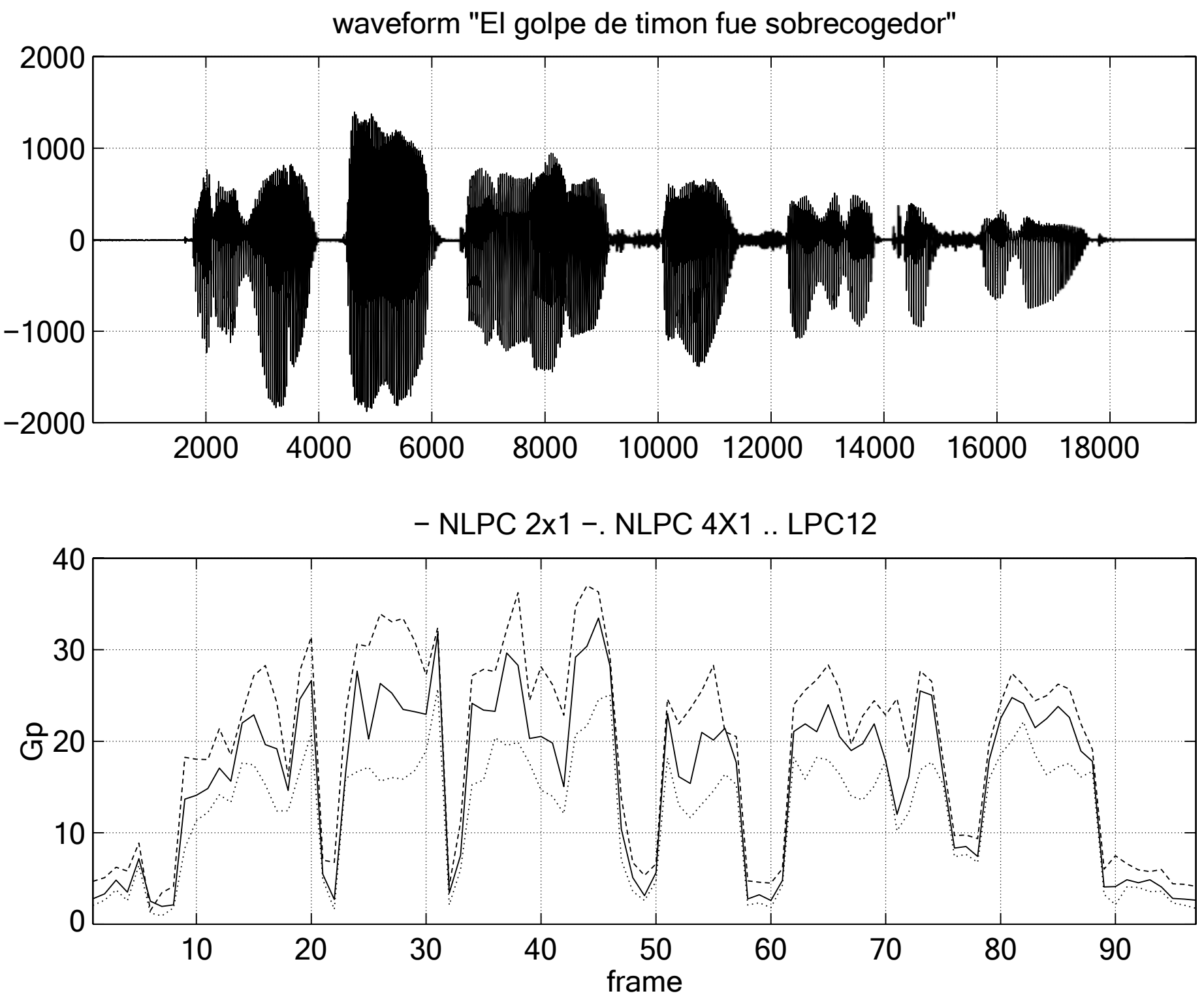

Fig. 1 Speech waveform and prediction gain of respectives frames. NLPC represents non-linear prediction with a MLP.

---

- Using four neurons in the hidden layer the prediction gain is significatively greater (see fig.1), but the number of parameters is doubled and also the computational burden.
- The training algorithm is the Levenberg-Marquardt, that is faster than back-propagation. Fig. 2 represents the sum-squared network error as function of epochs. Upper line is obtained with back-propagation, and the lower one with Levenberg-Marquardt, obtained with a typical voiced frame.

Fig. 3 represents the prediction gain for different weight's initializations, for the same frame. Upper line is MLP 4x1, lower one 2x1 and + is LPC-12The number of epochs has been fixed to 50, and 5 initializations per frame are realized. Figure 3 reveals that this predictor is very sensitive to the weights' initialization, so it is better to use 5 initializations and 50 epochs/initialization than 500 epochs and only one initialization

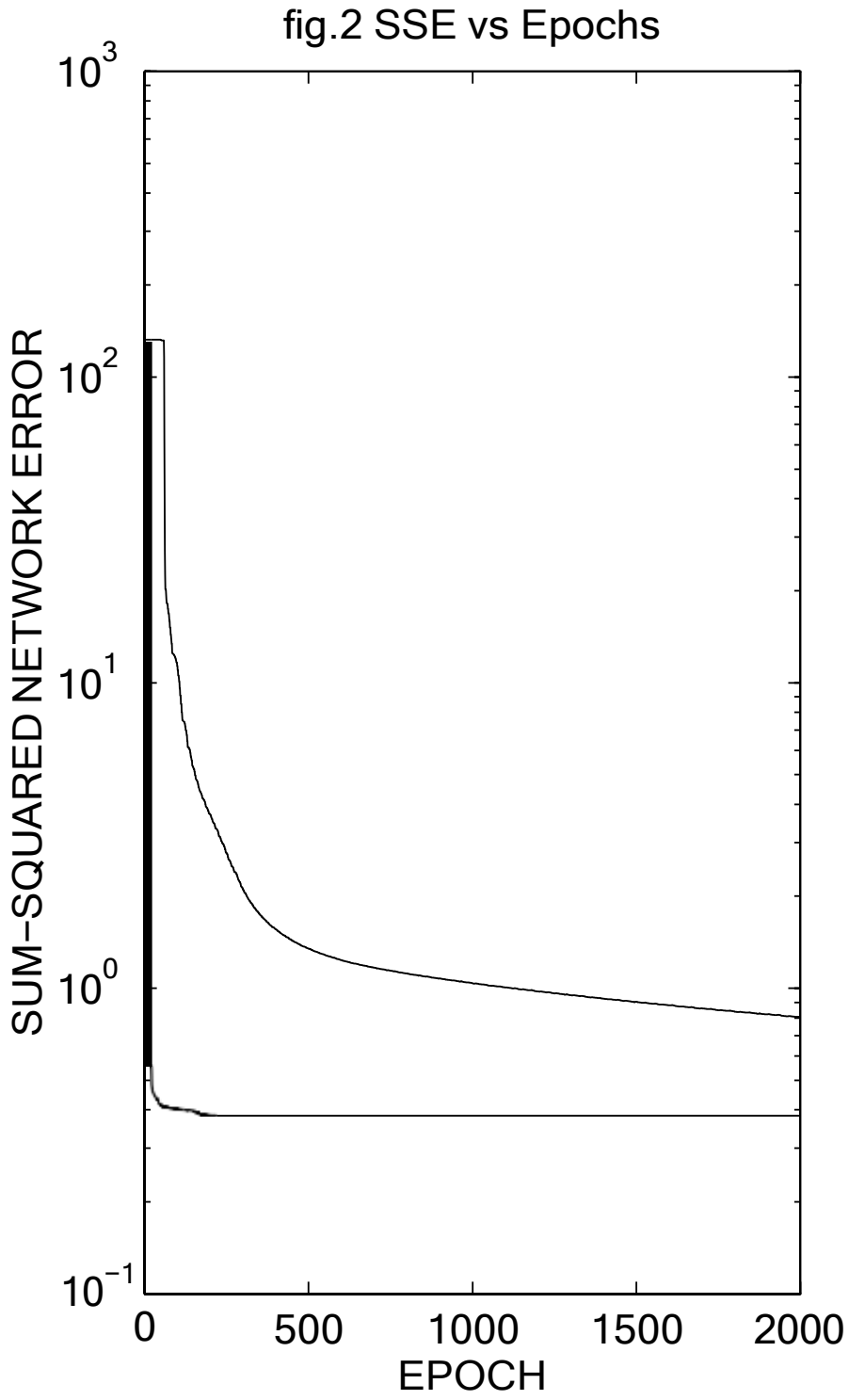

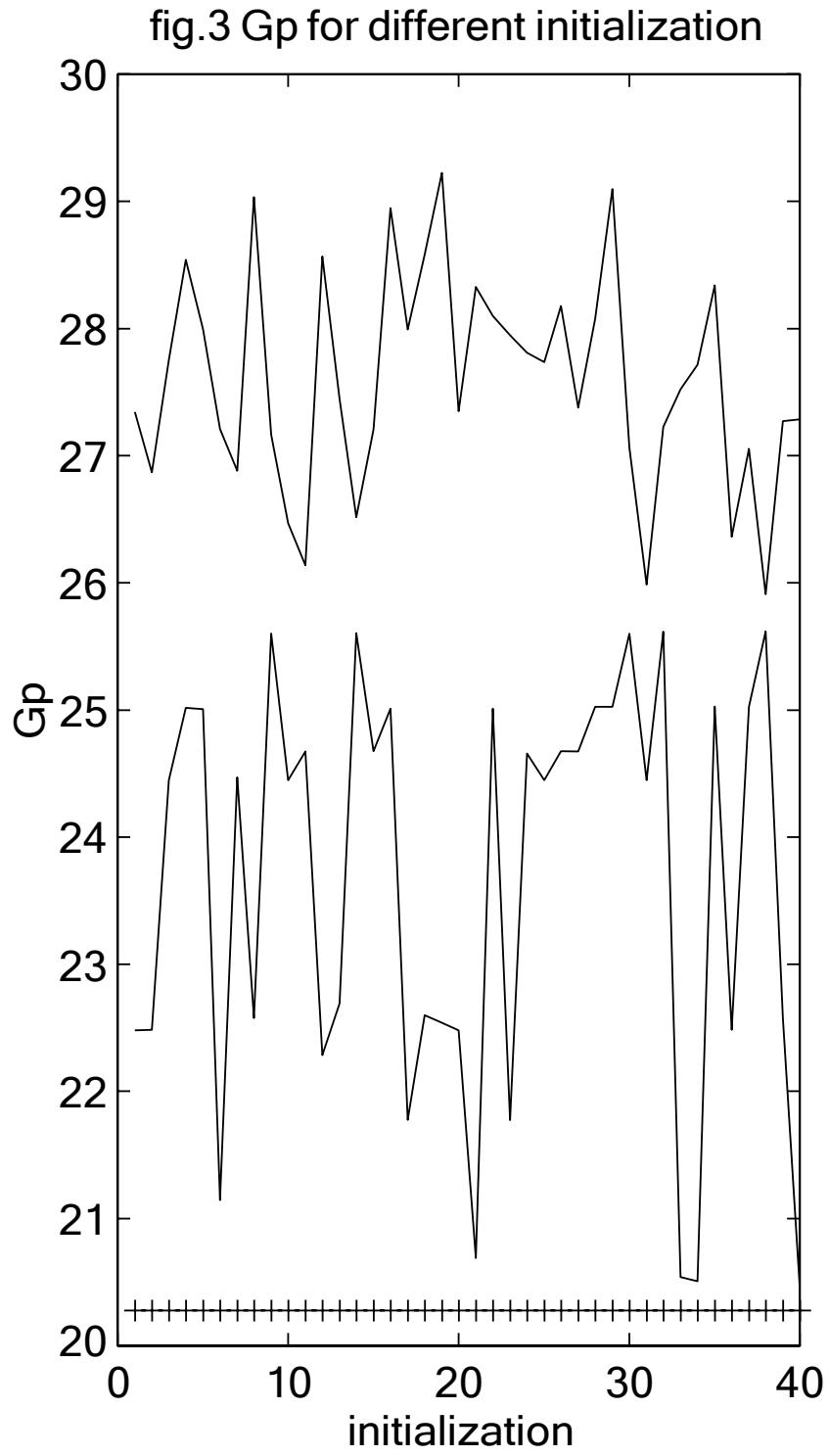

Fig. 2 SSE vs epochs. Upper line is bp, lower one is L-M. Fig. 3 represents Gp for different weigths' initialization. Upper line is MLP 4x1, lower one 2x1 and + is LPC-12

.

- The improvement obtained with the nonlinear predictor is greater for voiced portions of speech (see fig. 1) than for unvoiced segments, and greater for female speakers

In order to compare the performance of both predictors, we decided to plot in Ü$^2$ the pairs (linear prediction gain, nonlinear prediction gain), and the corresponding histograms of linear and nonlinear gains, as it can be seen in figure 4. Main observations are:

- Dots are located in the region y>x because Gnlpc >Glpc

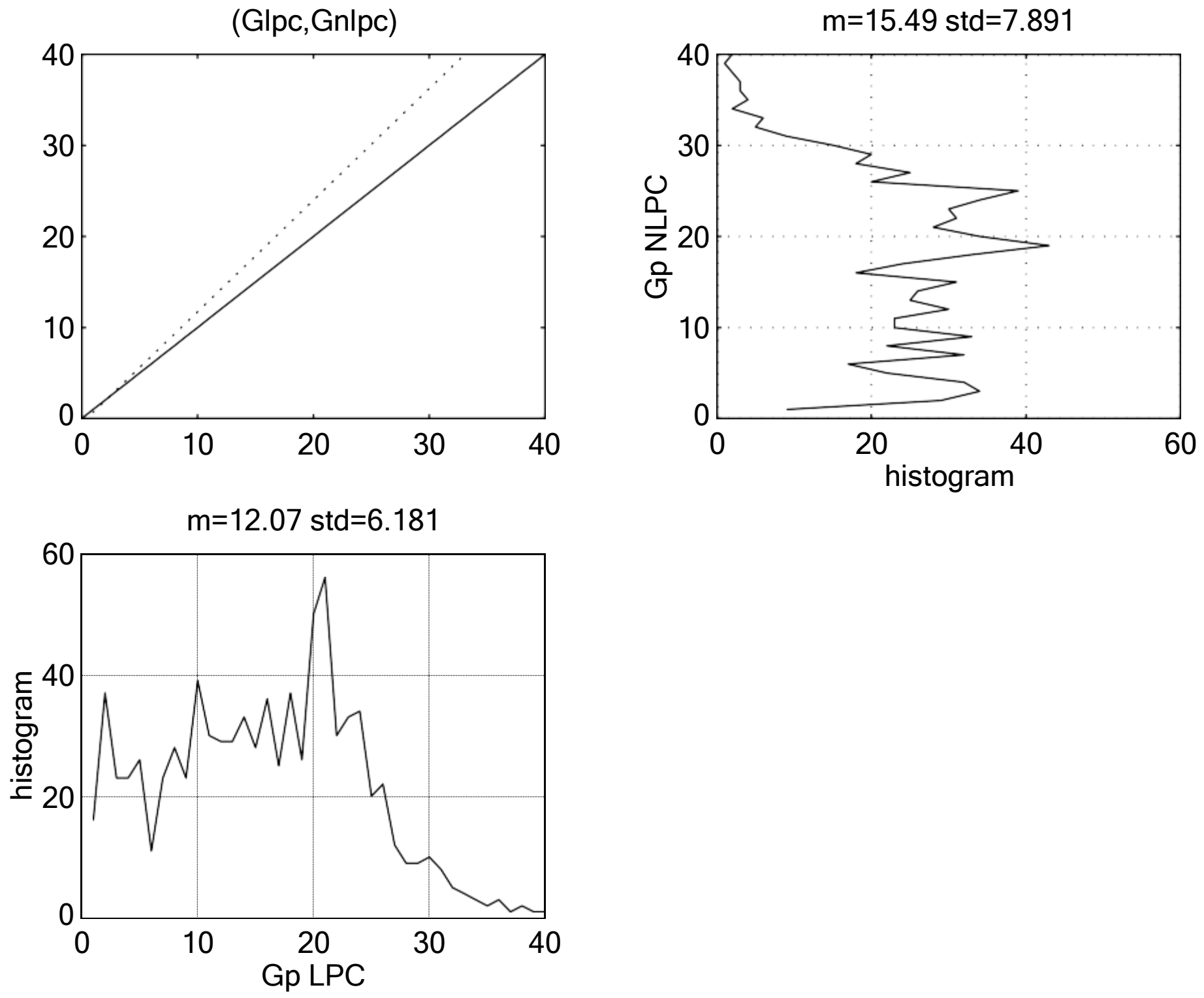

Figure 4 plot of Glpc vs Gnlpc

- The dash-doted line is fitted using linear regression
- There are two zones: the first one with low gains and variance corresponds to unvoiced frames. The second one with high gains and variance corresponds to voiced frames.

The main advantages of the nonlinear predictor with a neural net are:

- It is possible to adjust the prediction error easily than with linear prediction, because there are more parameters (in LPC the unique parameter is the predictionorder).
- It is possible to eliminate the pith periodicity of the residual error of voiced frames. Thus, it is not necessary a long term predictor, pitch measurement/transmission, etc [11].

- It is possible to integrate information. In linear prediction the input is the signal but with a neural net more information can be provided (energy, zero crossing rate, kind of frame, etc.)
- Although least-squares linear prediction will increase the gaussianity of the residual error, there are strong asymmetries that disapprove de claim of gaussianity for voiced sounds. Figure 5 shows the histograms of prediction error for a typical voiced frame. Top is LPC and bottom NLPC with a neural net. Cp and cn are counters for negative and positive sides of the histogram respectively. It can be seen that the residual error in the LPC is more asimmetrical than in the case of the neural net (NN).

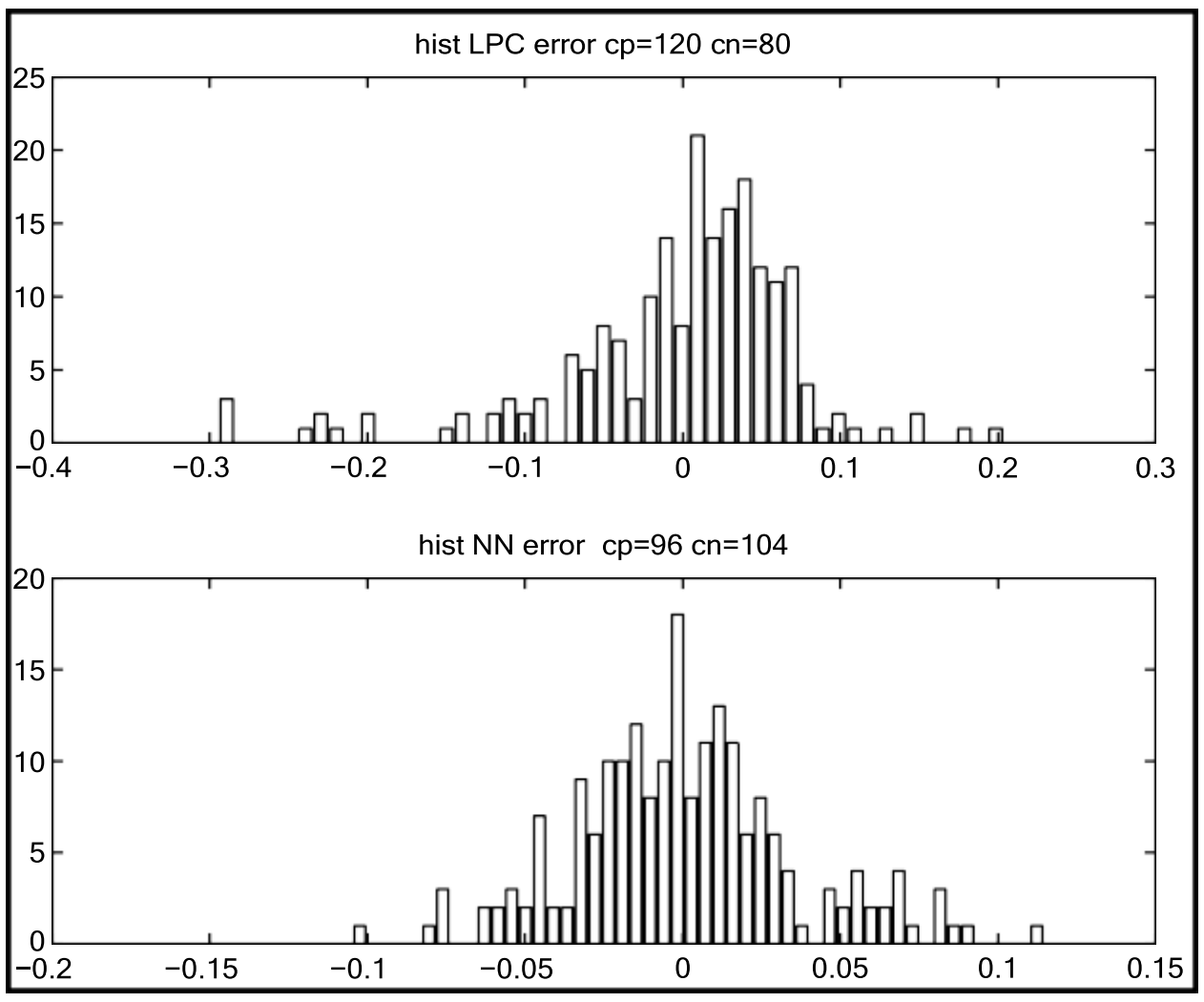

fig. 5 Histograms of prediction error

The main drawbacks of the nonlinear predictor are:

- Higher computational time. In a speech coding application this problem can be overcome with a codebook of predictors, because training is realized only one time.
- It is difficult to analyze a nonlinear system, because the superposition principle is not valid.

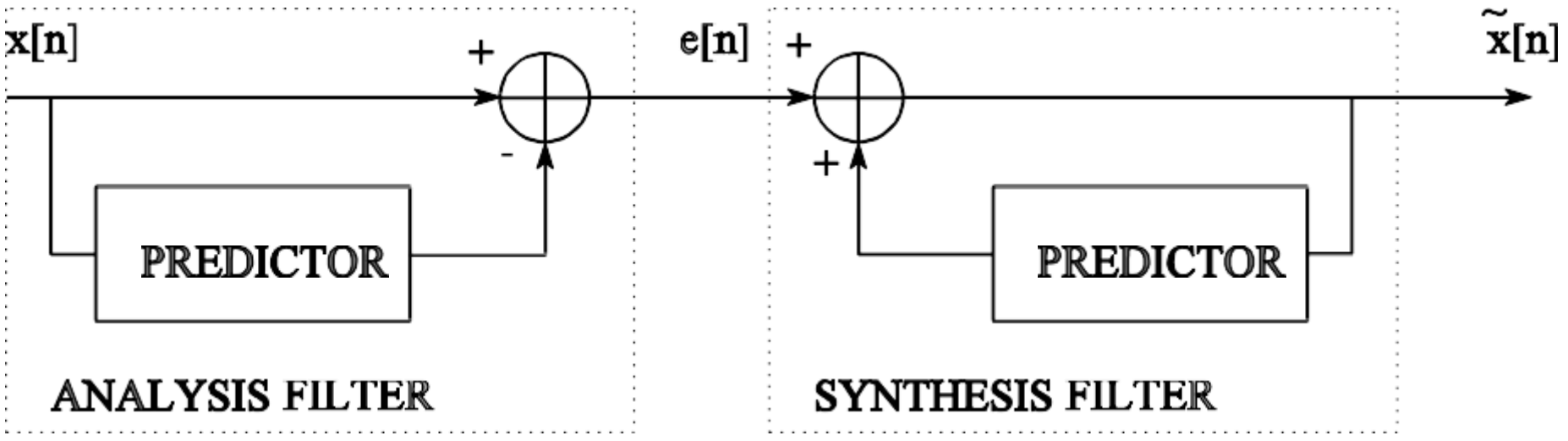

fig. 6 Analysis/synthesis structure

## 2.1 Parameters' quantization

In order to compare the effects of parameters' quantization in both structures (LPC and neural net), the degradation of prediction gain (Gp) is evaluated in twoexperiments:

a) Uniform quantization of the first layer NN and weights of the FIR LPC12 structure in the analysis configuration (see fig. 6). Figure 7b shows the evolution of the prediction gain as a function of the number of quantization bits, for the frame represented in figure 7a.

b) Evaluation of the prediction gain after analysis and synthesis with quantized parameters, in the same context of experiment a). Figure 7c shows the obtained results.

Main conclusions for analysis quantization are:

- In the analysis stage the LPC12 is better for a hard quantization, but for more than 7 bits the n. net outperforms LPC12.
- For a reasonable number of bits (9 bits) both methods obtain the same prediction gain than without quantization.

Main conclusions for analysis & Synthesis quantization are:

- NN outperforms LPC12. Specially, for a quantizer of few bits the LPC12 filter becomes unstable, degrading seriously the performance of the predictor.
- NN model obtains higher prediction gain because it captures the nonlinearities of the speech signal. Obviously, for a sufficient number of bits both methods reconstruct the original signal (except numerical errors), because analysis andsynthesis are inverse functions.

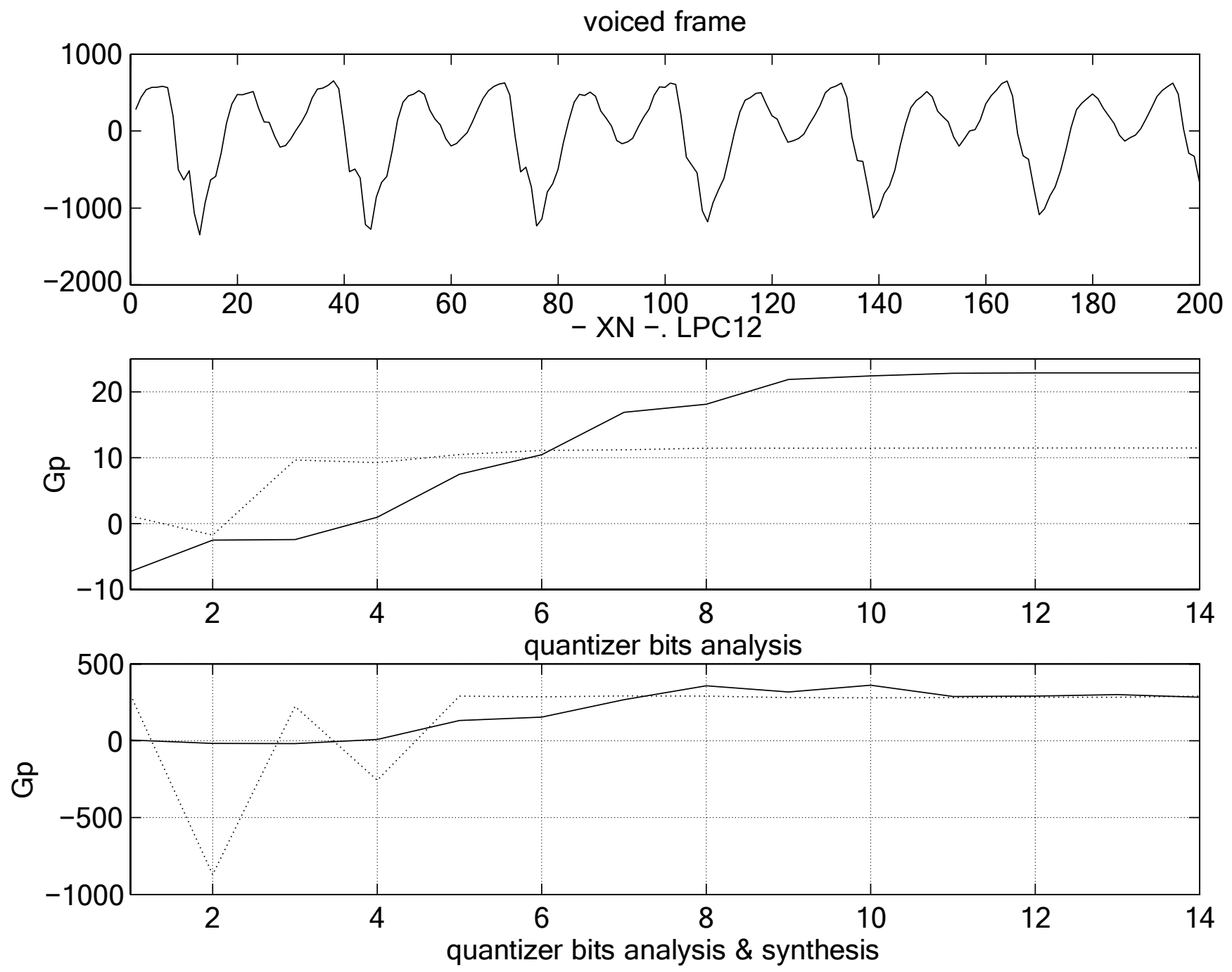

Fig. 7 voiced frame used for test and prediction gains obtained in analysis and analysis/synthesis.

Quantization of prediction error

If the prediction error (e[n] in fig. 6) is quantized, the residual error between the original signal and the output of the synthesis filter presents the following characteristics:

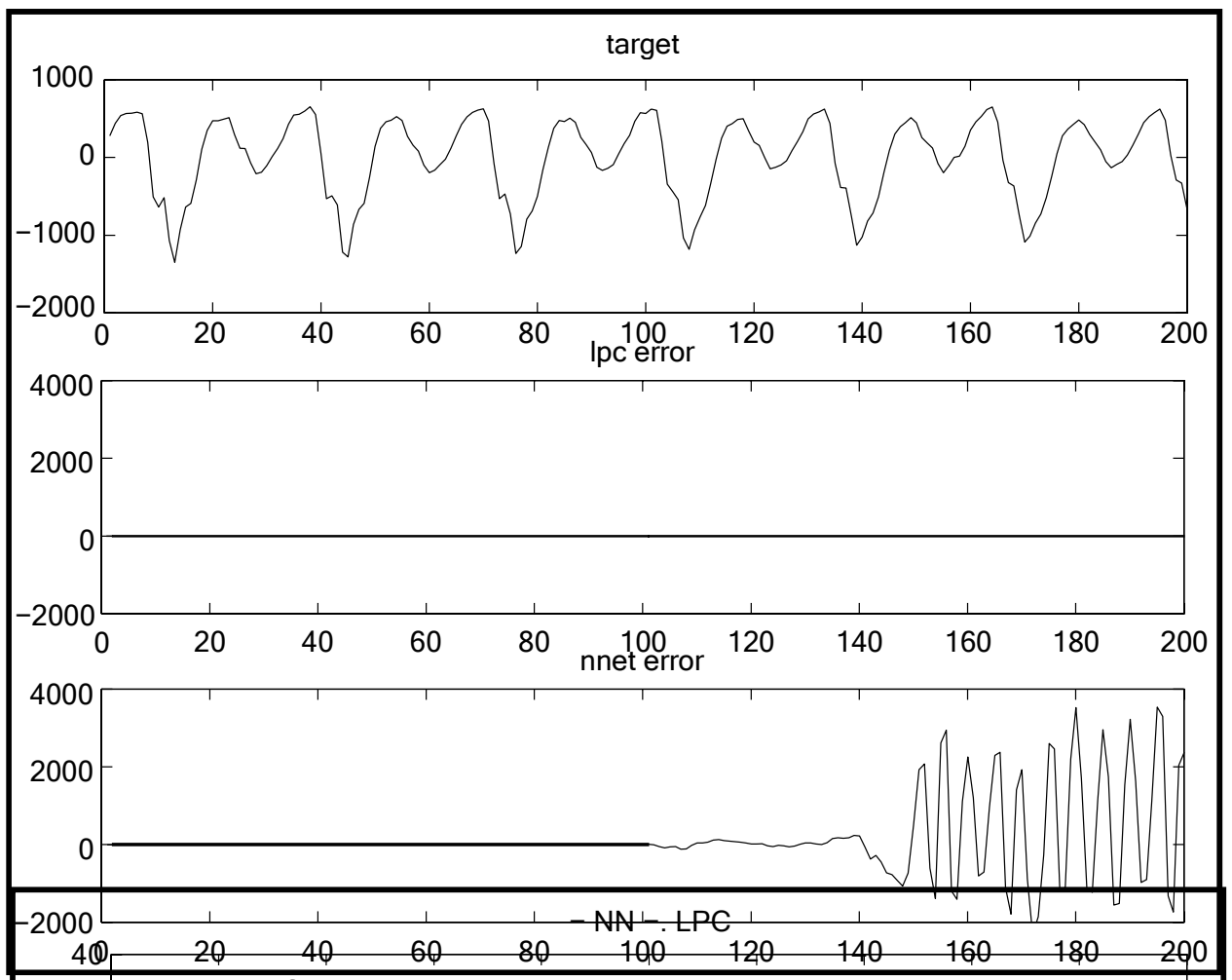

Fig. 8 error of reconstructed signal with error quantization

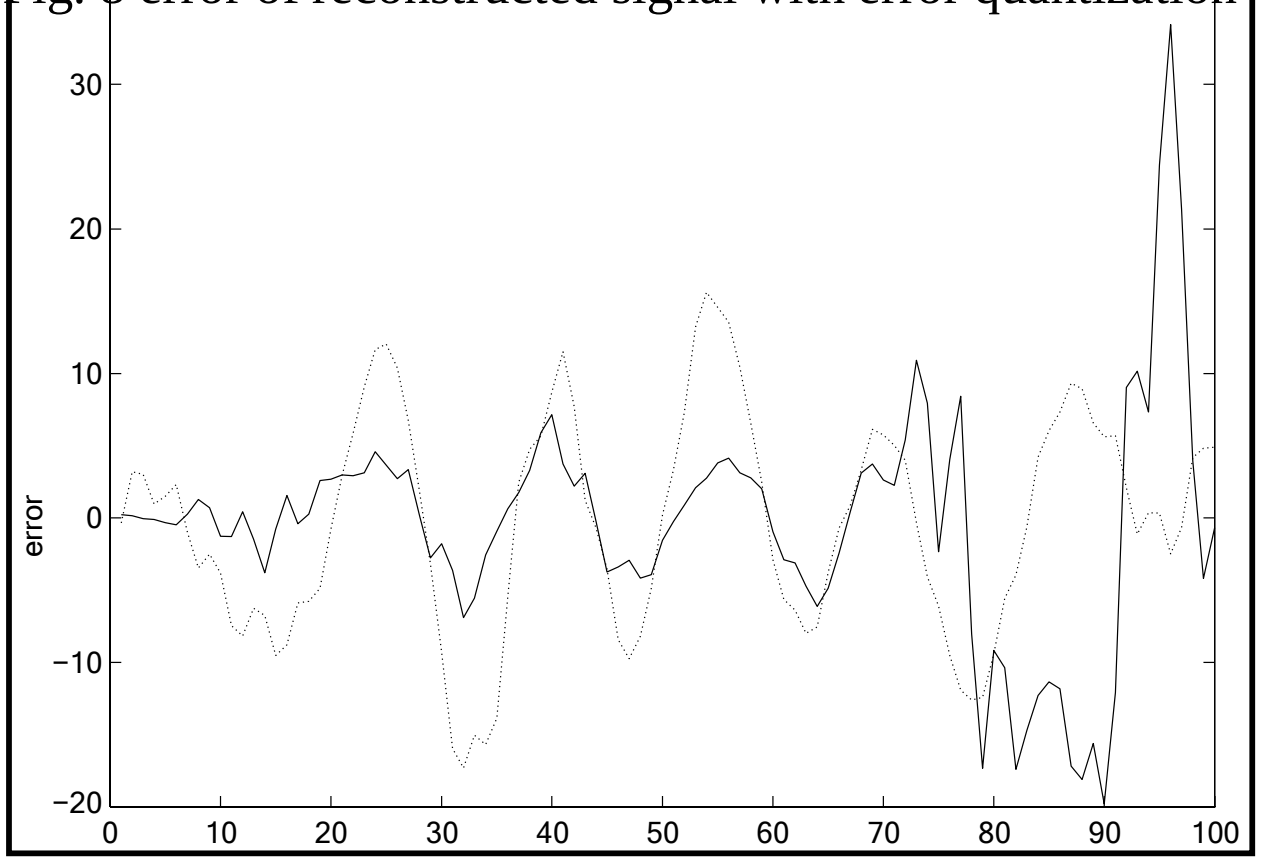

Fig.8b zoom of figure 8

For the nonlinear predictor there is an instant of time after which the filter becomes an oscillator. Figure 8 illustrates this phenomenon.

The useful duration of the nonlinear predictor is proportional to the number of quantizing bits. We think that this behaviour is due to the disagreement between training and testing conditions of the neural net. This problem, is not present with the linear predictor. Fig.8b is a zoom of the first samples. It shows that initially the error of the NN is smaller than LPC. The number of quantization bits in figures 8 and 8b is 9 bits.

Further study may be necessary for solving this problem. (Mainly, changing the way of training the network).

## 3. The lattice-MLP structure

With the parameters of last section, the neural net implements the following equation:

$$[n] \; w \; \hat{x}[n] \quad f_L \Big[ w_{31} \; f_{NL}\big(\sum_{i\;1}^{10} w_{1i} \; x[n \;\; i]\big) \quad b_{11}! \quad w_{32} \; f_{NL}\big(\sum_{i\;1}^{10} w_{2i} \; x[n \;\; i]\big) \quad b_{21}! \;\; b_2$$

where:

$$f_{NL}(a) \frac{1}{1 \quad e^{\;\; a}} \qquad ; \qquad f_L(a) \quad a$$

The last formula reveals that both summatories can be interpreted like a linear FIR filter. For an efficient implementation the delays (T) are shared by filters 1 and 2 (see fig. 9), but it is equivalent to the structure of figure 10. In this figure, the linear filters can be replaced with a lattice structure (fig.12) without modifying the implemented equation.

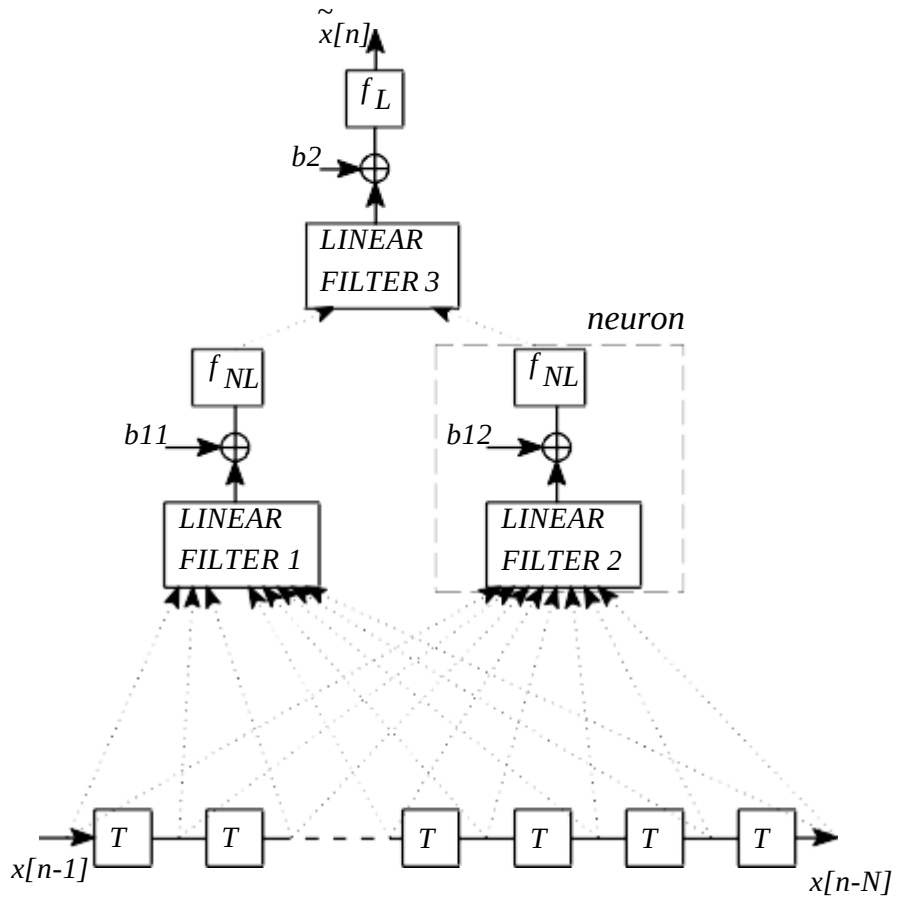

Fig.9 General structure for MLP predictor with minimum delays.

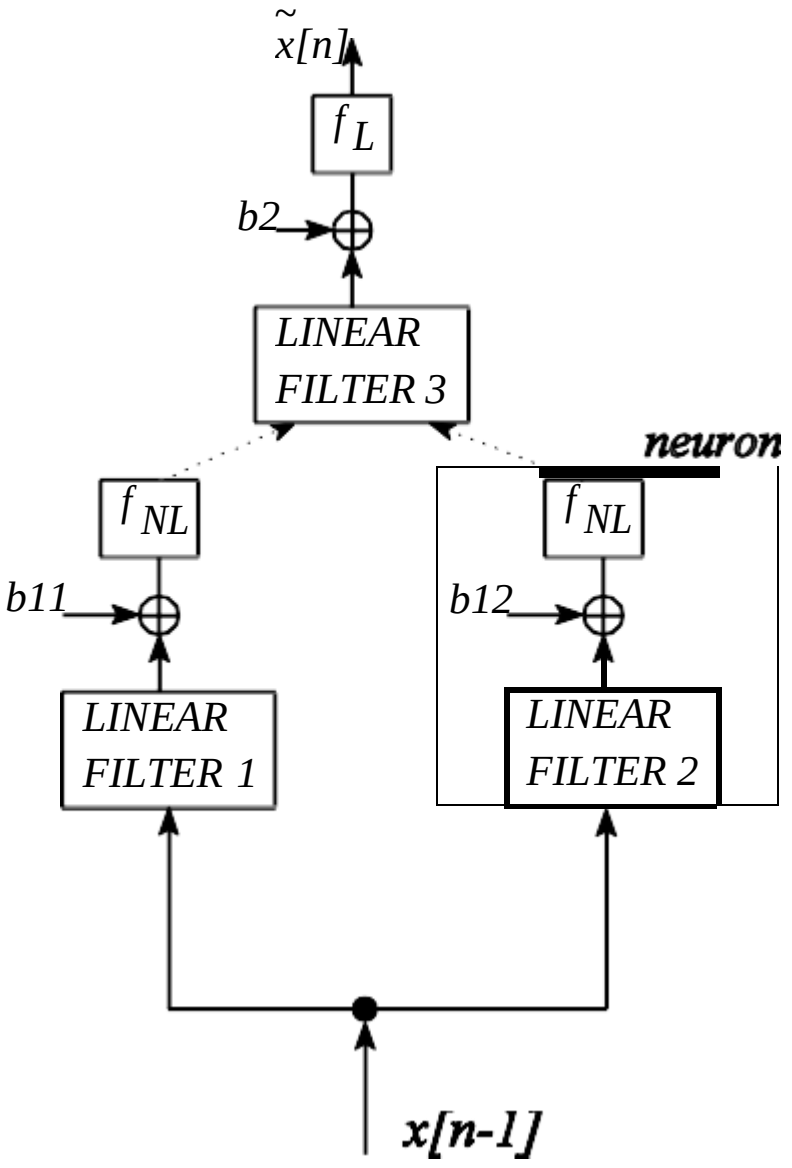

fig. 10 Structure of MLP predictor

The nonlinear predictor is the same of figure 9, but the linear filters can be implemented in the direct form (fig. 11) or with the lattice structure (fig. 12).

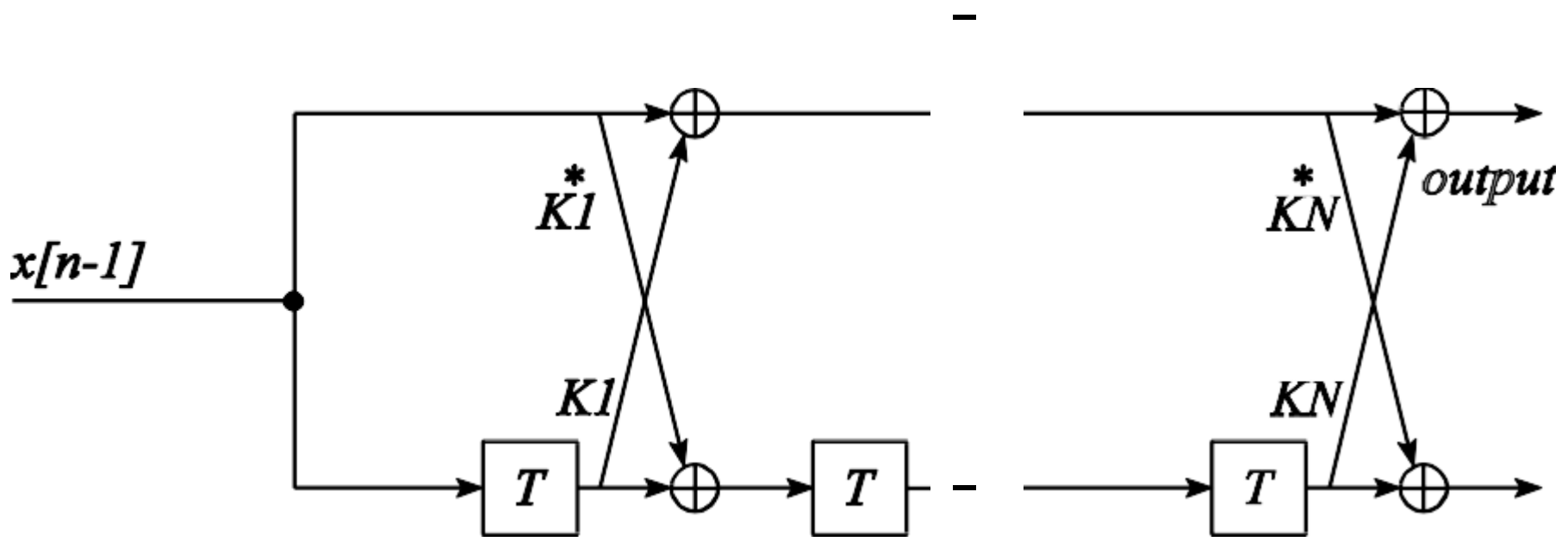

Fig. 12 lattice filter

Although the performance of both filters is identical without parameter quantization, the lattice-MLP presents the following advantages:

1. Its weights are better distributed, so they can be easily quantized. Figure 13 shows a logarithmic ratio between the standard deviation of FIR coefficients and the standard deviation of the lattice coefficients. It can be seen that the lattice coefficients behave better than the FIR coefficients.

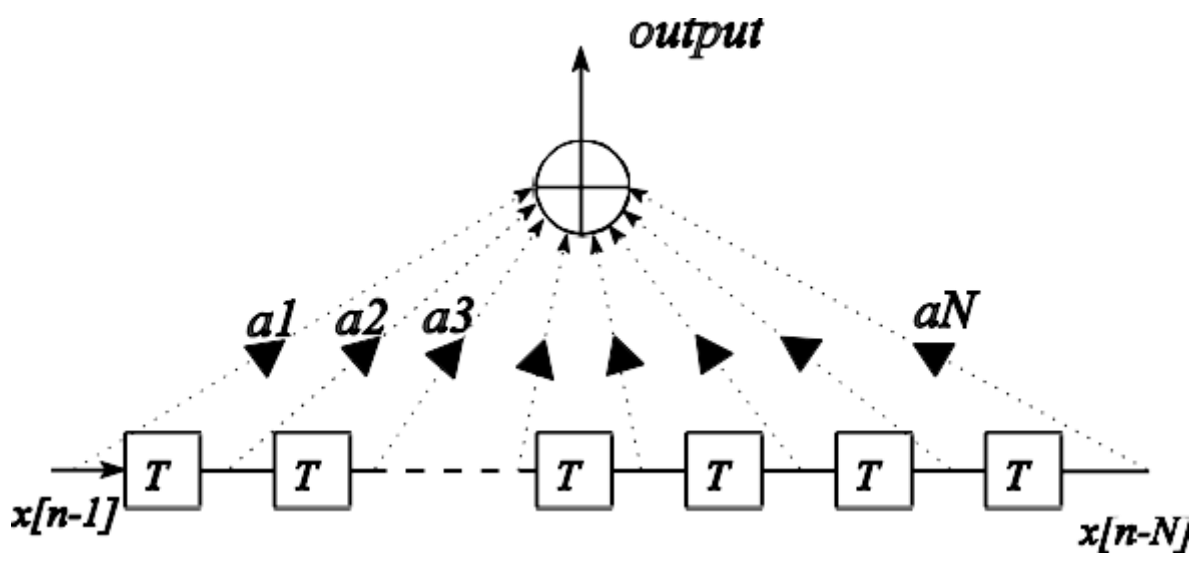

Figure 11. Structure of FIR filter.

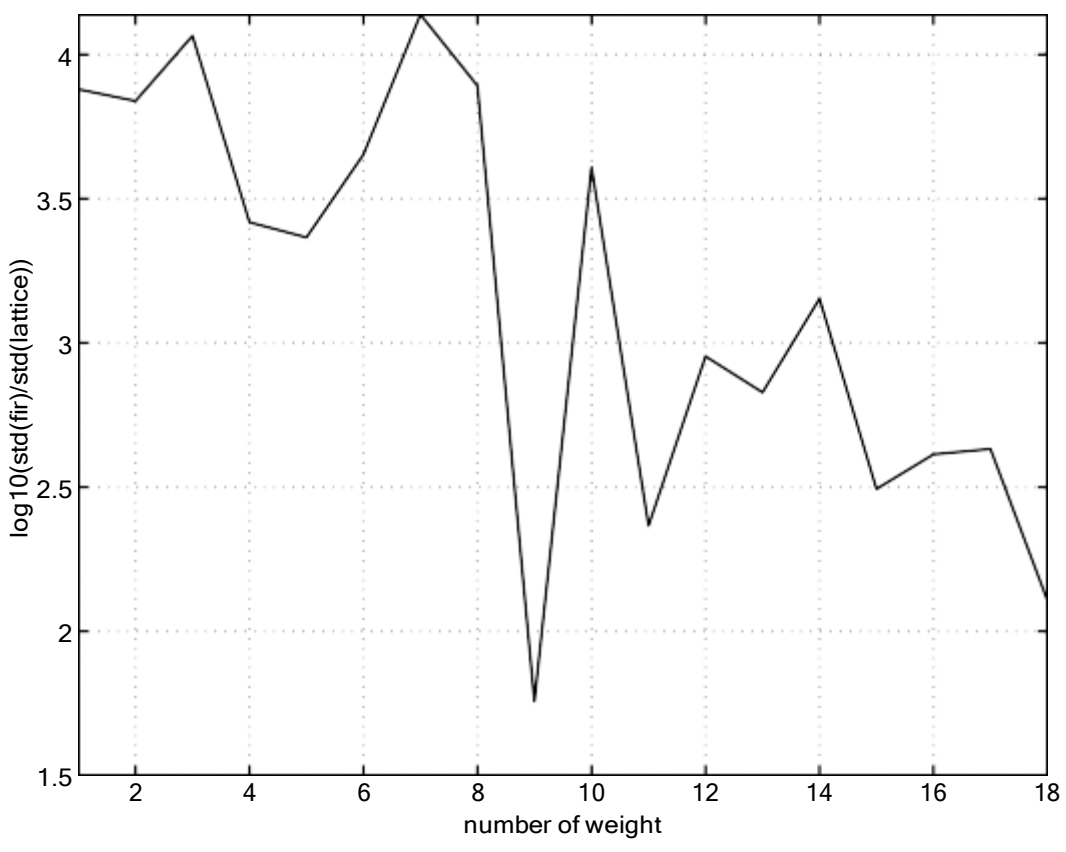

Fig. 13 logarithmic ratio σ(fir)/σ(lattice)

2. The parameters are ordered, so the first weights are more important than the final ones. So they can be better protected to transmission errors.

Preliminary results reveal the better distribution of lattice-MLP parameters (its variance is roughly 3700 times lower), and more robustness to transmission errors. Figure 14 represents the temporal evolution of the first weight (w1) in the classical MLP (dotted line) and the evolution of the first parcor (K1) with solid line, of the proposed lattice-MLP. High w1 values are truncated in order to see the amplitude of the small K1. This graphic reveals the smoother evolution and smaller dynamic range for the lattice-MLP, opening a great expectation for the achievable results with the proposed new structure.

## 4. Conclusions

This paper presents a comparison between prediction gain with a linear and nonlinear method. We have showed the superiority of the nonlinear predictor implemented with a MLP Several experiments have been done about parameters quantization of linear and nonlinear prediction models, revealing the robustness of the MLP.
In order to improve the performance of the nonlinear predictor, a modification of the traditional Multi-Layer Perceptron is proposed, based on the lattice filter.

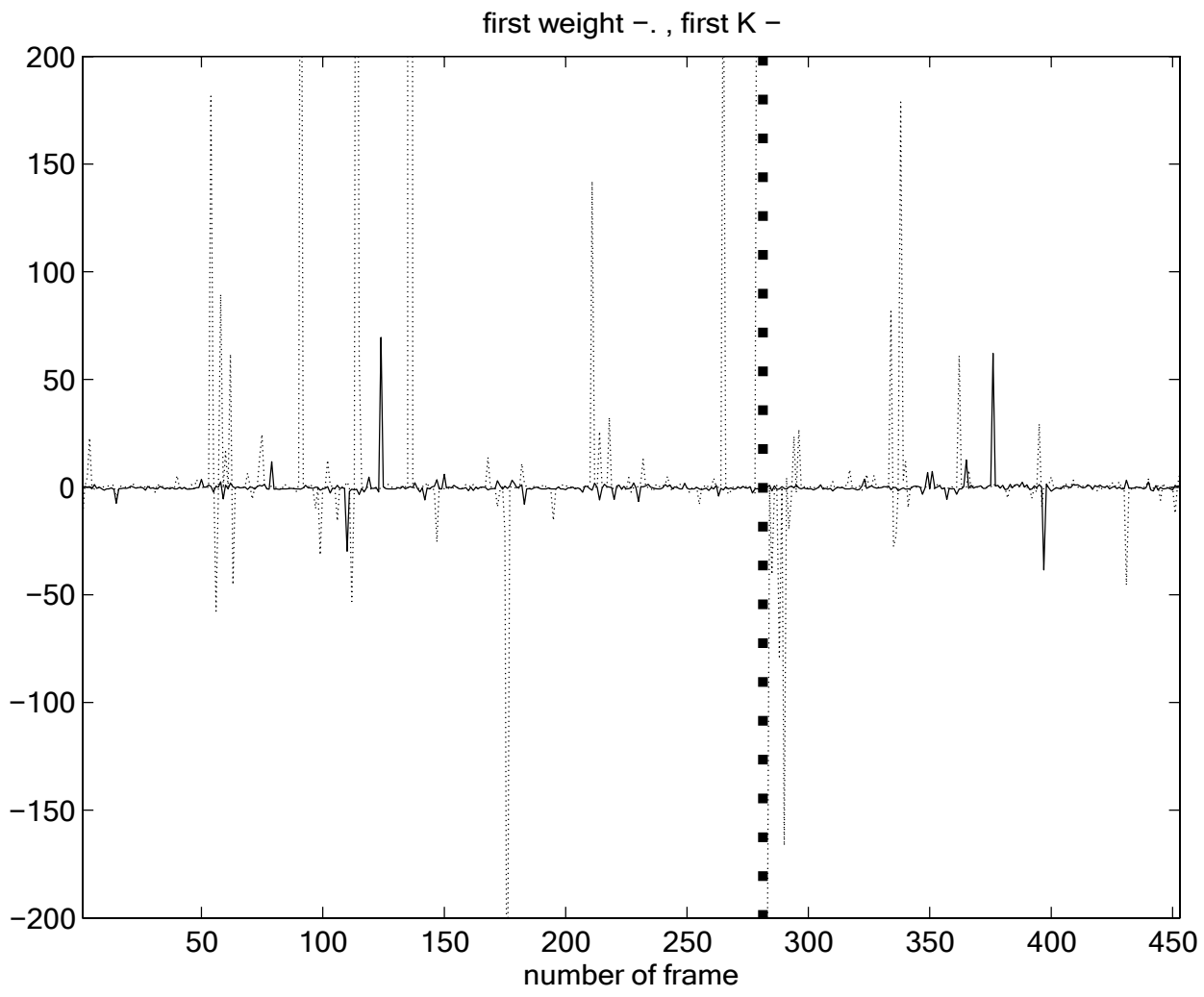

Fig. 14. Comparision between the first weight of the classical MLP and the first K of the lattice-MLP.